\documentclass[aps,twocolumn,showpacs,amsmath,amssymb,prl,superscriptaddress,floatfix]{revtex4-1}
\usepackage{graphicx}
\usepackage{bm}
\usepackage{bbold}
\usepackage{verbatim}
\usepackage{amsmath,amsfonts,amssymb,dsfont}
\usepackage{color}
\usepackage{hyperref}

\renewcommand{\c}{\mathbf{c}}
\newcommand{\F}{\mathbf{F}}

\renewcommand{\v}{\mathbf{v}}
\renewcommand{\L}{\mathbb{L}}
\newcommand{\X}{\mathbb{X}}
\newcommand{\T}{{\text{T}}}
\newcommand{\B}{\mathbf{B}}	
\newcommand{\Q}{\mathcal{Q}}
\newcommand{\calT}{\mathcal{T}}
\newcommand{\FN}{{\mathcal{F}}}

\newcommand{\FF}{{\mathcal{F}}}
\newcommand{\cN}{{\mathcal{C}_N}}
\newcommand{\cNsq}{{\mathcal{C}_N^2}}
\newcommand{\D}{\mathbb{D}}
\newcommand{\J}{\mathbf{J}}
\newcommand{\Y}{\mathbb{Y}}
\newcommand{\f}{\hat{f}}
\newcommand{\TC}{\mathcal{T}}

\definecolor{green}{rgb}{0.0, 0.5, 0.0}

\newcommand{\er}[1]{\eqref{#1}}

\begin{document}

\title{Unified Thermodynamic Uncertainty Relations in Linear Response}

\author{Katarzyna Macieszczak}
\affiliation{School of Physics and Astronomy, University of Nottingham, University Park, Nottingham NG7 2RD, United Kingdom} 
\affiliation{Centre for the Mathematics and Theoretical Physics of Quantum nonequilibrium Systems, University of Nottingham, University Park, Nottingham NG7 2RD, United Kingdom}
\affiliation{TCM Group, Cavendish Laboratory, University of Cambridge, J. J. Thomson Ave., Cambridge CB3 0HE, United Kingdom}
\author{Kay Brandner}
\affiliation{Department of Applied Physics, Aalto University,
00076 Aalto, Finland}
\author{Juan P.\ Garrahan}
\affiliation{School of Physics and Astronomy, University of Nottingham, University Park, Nottingham NG7 2RD, United Kingdom} 
\affiliation{Centre for the Mathematics and Theoretical Physics of Quantum nonequilibrium Systems, University of Nottingham, University Park, Nottingham  NG7 2RD, United Kingdom}


\date{\today}

\begin{abstract}
Thermodynamic uncertainty relations (TURs) are recently established relations between the relative uncertainty of 
time-integrated currents and entropy production in nonequilibrium
systems. 
For small perturbations away from equilibrium, linear response (LR)
theory provides the natural framework to study generic nonequilibrium
processes. 
Here we use LR to derive TURs in a straightforward and unified way.
Our approach allows us to generalize TURs to systems without local 
time reversal symmetry, including, for example, ballistic transport, and
periodically driven classical and quantum systems. 
We find that for broken time reversal, the bounds on the relative
uncertainty are controlled both by dissipation and by a parameter
encoding the asymmetry of the  Onsager matrix. 
We illustrate our results with an example from mesoscopic physics. 
We also extend our approach beyond linear response: for Markovian
dynamics it reveals a connection between the TUR and current fluctuation theorems.
\end{abstract}

\maketitle

\smallskip
{\em Introduction}.--- 
Central to modern statistical mechanics are general principles 
governing the behavior of fluctuations in systems away from thermal
equilibrium. 
The simplest of these principles is the connection between the change
of expectation values of observables in response to small 
perturbations and correlations of spontaneous fluctuations in 
equilibrium, the fluctuation-dissipation theorem (FDT) \cite{Kubo2012}. 
For systems arbitrarily far from equilibrium, fluctuation theorems
\cite{Gallavotti1995,Jarzynski1997,Crooks1999,Lebowitz1999} provide 
the most general characterization to date of the statistical 
properties of fluctuations. 
These general principles are not only of fundamental and conceptual
importance, but also of practical benefit as they connect the 
hard-to-compute fluctuations in a specific system with the easier
accessible constraints determined by general properties such as
symmetry. 
For example, FDT is exploited to obtain transport coefficients from
equilibrium time-correlators via Green-Kubo relations
\cite{Green1954,Kubo1957}, and equilibrium free-energy differences 
can be recovered from nonequilibrium trajectories via the 
Jarzynski relation \cite{Jarzynski1997}.

An important recent addition to the above has been the discovery
of general lower bounds on the fluctuations of time-integrated 
currents in nonequilibrium steady states \cite{Barato2015,
Gingrich2016,Polettini2016,Pietzonka2016,Pietzonka2016b,Gingrich2016b,
Seifert2017} of stochastic systems. 
In particular, for Markovian dynamics with local detailed balance, 
given a time-integrated current $J_\alpha(t)$, whose long-time average
converges to $ \langle J_\alpha(t) \rangle/t \to J_\alpha \neq 0$, and
variance, $\left[\langle J_\alpha(t)^2\rangle-\langle J_\alpha(t)
\rangle^{2} \right]/t$, to $D_\alpha\neq 0$, the {\em thermodynamic
uncertainty relation} (TUR) \cite{Barato2015} provides a general 
constraint: the squared relative uncertainty, $\varepsilon^2(t)
=\left[\langle J_\alpha(t)^2\rangle-\langle J_\alpha(t) \rangle^{2}
\right]/\langle J_\alpha(t)\rangle^2$, asymptotically obeys the
inequality  \cite{Barato2015,Gingrich2016} 
\begin{equation}\label{TUR}
\varepsilon^2(t) \,\sigma t \to  \sigma D_\alpha / J_\alpha^2
\geq 2,
\end{equation}
where $\sigma$ is the rate of entropy production. 
This bound implies that more precise output (smaller $\varepsilon$),
requires more dissipation $\sigma t$. 
The TUR \er{TUR} pertains to small deviations around the average
\cite{Barato2015,Pietzonka2016}, but was shown \cite{Gingrich2016} to follow, for 
time homogeneous Markov processes, from a general bound also valid in
the large deviation regime. 
Both TURs and bounds on large deviation functions have been refined 
and extended \cite{Polettini2016,Pietzonka2016b,Gingrich2016b,Seifert2017,Pigolotti2017}, adapted to
counting observables \cite{Garrahan2017}, to first-passage times
\cite{Garrahan2017,Gingrich2017}, generalized to finite times 
\cite{Pietzonka2017,Horowitz2017,Manikandan2017}, to discrete time and
periodic dynamics \cite{Proesmans2017,Chiuchiu2017,Barato2018}, and applied to a
variety of nonequilibrium problems \cite{Barato2015b,Pietzonka2016c,
Hyeon2017,Hwang2017,Brandner2017,Rosas2017,Fischer2017}.

In this Letter, we consider TURs from the general point of view of 
{\em linear response} (LR) as applicable to systems where a
nonequilibrium state (steady or periodic) arises due to small 
perturbations. 
In this regime, linear irreversible thermodynamics applies 
\cite{Callen1985}: 
a small stationary current $J_\alpha$, e.g. a particle or 
heat current, can be expanded in terms of {\em affinities} $F_\alpha$, such as chemical potential or temperature differences, as $J_\alpha=
\sum_\beta L_{\alpha\beta} F_\beta$, where the response coefficients
$L_{\alpha\beta}$ form the {\em Onsager matrix} $\L$.
Within this framework, the FDT implies $\partial J_\alpha / \partial
F_{\alpha} = D_\alpha/2$, with $D_\alpha=2 L_{\alpha\alpha}$ 
describing Gaussian fluctuations near equilibrium, while the average
rate of entropy production is $\sigma=\sum_{\alpha}F_\alpha J_\alpha$
(also valid beyond LR). 
The strength of LR is that it can be applied irrespective of whether
the perturbed system obeys local time reversibility, with the relevant
features of the dynamics encoded in the Onsager matrix. 
Thus, it can be used to describe ballistic transport in a magnetic
field, periodically driven systems \cite{Brandner2015}, and open 
quantum systems close to equilibrium \cite{Brandner2016b}. 

Here, we show that, within LR, TURs can be derived in a 
straightforward and unified manner that accounts for systems with
generic dynamical properties. 
In particular, we find that for any current, i.e., for any contraction
of basis currents  $J_\c=\sum_\alpha c_\alpha J_\alpha$, the general 
TUR
\begin{equation}\label{result} 
\sigma D_\c / J_\c^2 \geq 2 / \left( 1+s_{\L}^2 \right)
\end{equation}
holds.
Here, $s_{\L}$ is the {\em asymmetry index} of the Onsager matrix
\cite{Crouzeix2003,Brandner2013}, which quantifies the extent to which
the breaking of time-reversal symmetry affects response properties. 
We will illustrate this general TUR \er{result} by discussing chiral 
transport in a mesoscopic multiterminal conductor
\cite{Buttiker1992,Sivan1986,Baranger1989,Butcher1990,Matthews2014}. 

Extending our approach beyond LR, we introduce a variational principle
that allows us to find the current with the smallest 
uncertainty. 
In the time-reversible case, this makes it possible to establish
a connection between the TUR \eqref{TUR} and fluctuation theorems
\cite{Gallavotti1995,Gallavotti1996,Andrieux2007,Barato2015c,Ray2017}.
We also discuss generalized TURs for chiral transport beyond LR.

{\em Linear response bounds}.---
Consider measuring a current $J_\c$ given by a linear combination of 
basis currents, $J_\c=\sum_\alpha c_\alpha J_\alpha = \c^\T 
\L\F$, where $\c$ is a vector of real coefficients, 
$(\c)_{\alpha}=c_\alpha$, and $\F$ is a vector of affinities, 
$(\F)_\alpha = F_\alpha$.
In LR, the fluctuations of this current around the stationary value 
$J_\c$ are given by $D_\c=2\sum_{\alpha\beta}c_\alpha L_{\alpha\beta} 
c_\beta=2\c^{\T}\L\c$, as $\L$ also describes the correlations between
Gaussian fluctuations of the basis currents \cite{Callen1985}. 
Its relative precision (inverse of the relative uncertainty) is 
bounded from above by that of the current with the lowest relative
uncertainty, 
\begin{equation}\label{prec}
\frac{J_\c^2}{\sigma D_\c} 
\leq\max_\c \frac{J_\c^2}{\sigma D_\c}=
\max_\c\frac{(\c^{\T}\L\F)^2}{2\F^{\T}\L\F\,\c^{\T}\L\c},
\end{equation}
where we have included  the rate of entropy production $\sigma=
\sum_{\alpha}F_\alpha J_\alpha=\F^{\T}\L\F$ in the denominator
\footnote{Note that the trivial lower bound on the relative precision,
$0 \leq J_\c^2/\sigma D_\c$, is attained whenever $\c$ is orthogonal 
to $\L\F$ such that $J_\c=0$.}. 

For time-reversal symmetric systems, the Onsager matrix is symmetric
\cite{Callen1985}. 
In general, however, we have $\L =\L_S +\L_A$, where $\L_S=\L_S^\T$ is
the symmetric and $\L_A = -\L_A^\T$ the antisymmetric part of $\L$. 
For any real coefficients $\c$, we have that current fluctuations are
determined only by the symmetric part of 
$\L$, $\c^\T\L\c = \c^\T\L_S\c$, which, thus, must be positive 
semidefinite. 
This condition is also implied by the second law \cite{Seifert2012}, 
as $\sigma=\F^{\T}\L\F \geq 0$. 

{\em (i) Time-reversible case}:
First, we consider systems with a symmetric Onsager matrix, $\L=\L_S$,
such as time-homogeneous Markov processes with local detailed balance.
The numerator in \er{prec} can then be written as the square of the
scalar product of $\L^{1/2}\c$ and $\L^{1/2} \F$. 
Using the Cauchy-Schwarz inequality, $(\c^{\T}\L\F)^2 \leq (\c^{\T}\L
\c)(\F^{\T}\L\F)$, we obtain the time-symmetric TUR
\begin{equation}\label{max}
J_\c^2/(\sigma D_\c)\leq 1/2.
\end{equation}
Note that \er{max} is saturated if  $\L^{1/2}\c\parallel\L^{1/2}\F$. 
This condition requires $\c\parallel\F$ on the orthogonal complement 
of the kernel of $\L$, where $\L^{1/2}$ can be inverted. 
In particular, for positive $\L$, the only current saturating the 
inequality is proportional to the affinity vector $\F$, i.e., the 
entropy production \cite{Pietzonka2016}.  
For this choice of current
in local detail balance dynamics, the quadratic bound on the rate function by the entropy production is also the tightest \cite{Gingrich2016,
Gingrich2016b,Horowitz2017,Nardini2018}.
Notably, for $\c$ chosen as the $\nu$-th eigenvector of the Onsager
matrix, $\L\c = \lambda_\nu\c$, we obtain the even stronger equality
\begin{equation}
J_\c^2/D_\c= \lambda_\nu F_\nu^2 /2,
\label{strong}
\end{equation}
which involves only the entropy production rate along the $\nu$th
direction as $\sigma=\sum_{\nu}\lambda_{\nu} F_{\nu}^2$ in the 
diagonal basis of $\L$, see, also, \cite{Gingrich2016}. 

{\em (ii) Time-nonreversible case}: 
Assuming that $\L_S$ is positive and, thus, invertible, we consider the
numerator in \er{prec} as the square of the scalar product of 
$\L_S^{1/2} \c$ and $\L_S^{-1/2} \L \F$. 
Via the Cauchy-Schwarz inequality we obtain
\begin{equation}\label{maxA}
\frac{J_\c^2}{\sigma D_\c} \leq
\frac{\F^{\T}\L^{\T}\L_S^{-1}\L\F}
{2\F^{\T}\L_S\F}=\frac{1}{2} + \frac{\F^{\T}\L_A^{\T}\L_S^{-1}\L_A\F}
{2 \F^{\T}\L_S\F}
.
\end{equation}
This inequality is saturated for 
\begin{equation}\label{optC_LR}
\c_\text{opt} \, \propto\, \L_S^{-1}\L \F= \F+ \L_S^{-1}\L_A \F, 
\end{equation}
which is generally \emph{not parallel} to the affinity vector $\F$, as a consequence of the average currents 
being determined by the full $\L$, while the current fluctuations 
depend only on $\L_S$. Since the choice $\c \parallel \F$ as in \er{max}, i.e., the entropy rate current, gives $J_\F^2/(\sigma D_\F) =1/2$, cf.  \er{max}, the last term in \eqref{maxA} is necessarily positive and
the inequality is \emph{weaker} than in the symmetric case. This manifests  the existence of reversible currents $J_\alpha^\text{rev}=
(\L_A\F)_\alpha$, which, in contrast to the irreversible currents, 
$J_\alpha^\text{irrev}=(\L_S\F)_\alpha$, do not contribute to the 
total rate of entropy production or the variance of a current~\cite{Brandner2013b,Brandner2013}, thus giving rise to more precise currents $J_\c$ that exceed the time-reversible bound \er{max}. Furthermore,~\eqref{optC_LR} and,  thus, the value of rhs of~\eqref{maxA}, can be determined from long-time averages, $\langle J_\alpha(t)\rangle/t \to (\L \F)_\alpha$,
 and equal-time correlations, $[\langle J_\alpha(t)J_\beta(t)
 \rangle-\langle J_\alpha(t) \rangle \langle J_\beta(t) \rangle]/t
 \to 2(\L_S)_{\alpha\beta}$ without the need to vary the affinities, as required
 to recover $\L$~\cite{Callen1985}.

The bound \eqref{maxA}, depends on affinities, which, in principle, can
be tuned in an experimental setup. 
The \emph{fundamental bound} on current uncertainty, which is
independent from affinities, is given by
\begin{align}
&
\frac{J_\c^2}{\sigma D_\c}  \leq
\frac{1}{2} +\max_\F\frac{\F^{\T}\L_A^{\T}\L_S^{-1}\L_A\F}{
2 \F^{\T}\L_S\F}
\label{maxF} \\
& \;\;\;\;\;\;\;\;
= \frac{1}{2}  + 
\max_{\tilde{\F}}
\frac{\tilde{\F}^{\T} \L_S^{-1/2} \L_A^{\T} \L_S^{-1} \L_A \L_S^{-1/2}
\tilde{\F}}{2 \tilde{\F}^{\T}\!\tilde{\F}}
= \frac{1+s_{\L}^2}{2}, \nonumber 
\end{align}
where $\tilde{\F}=\L_S^{1/2} \F$, and 
\begin{equation}\label{sL}
s_{\L}= \big\lVert \L_S^{-1/2} (i\L_A) \L_S^{-1/2} \big\rVert 
\end{equation}
is the maximal eigenvalue of the (asymmetric) Hermitian matrix $\L_S^{-1/2} (i\L_A) \L_S^{-1/2}=\X$ [where 
$\X^\dagger \X=\L_S^{-1/2} \L_A^{\T} \L_S^{-1}\L_A \L_S^{-1/2}$ appears in the second line of ~\eqref{maxF}]. Therefore, in order to saturate \eqref{maxF}, the affinities must be chosen 
as $\F_\text{opt}=\L_S^{-1/2} \tilde \F_\text{opt}$ with 
$\tilde \F_\text{opt}$ belonging to the double-degenerate $s_\L^2$ eigenspace of $\X^\dagger \X$~\footnote{The corresponding optimal current is
	$\c_\text{opt}=\F_\text{opt}+\L_S^{-1}\L_A \F_\text{opt}
	=\L_S^{-1/2}(1+\X) \F_\text{opt}
	=\L_S^{-1/2}\tilde{\c}_\text{opt}$. Here $\tilde{\c}_\text{opt}$ is also a  $s_\L^2$-eigenvector of $\X^\dagger \X$, since the spectrum of $\X$ is reflective with respect to\ $0$, i.e.,  
	pairs of eigenvectors with opposite eigenvalues are connected by complex
	conjugation.}.

The parameter $s_{\L}$ is known as the \emph{asymmetry index} of the
Onsager matrix $\L$, i.e., the minimal value of 
$s$ such that $s \, \L_S + i\L_A$ is nonnegative over complex 
vectors \cite{Crouzeix2003,Brandner2013}.  Since $s_{\L}$ depends on the Onsager matrix $\L$,  
the bound \eqref{maxF} [or \eqref{result}] is no longer strictly universal, in contrast to the time-reversible one~\eqref{max}. It is important to note that our result~\eqref{maxF}, however, still implies a semiuniversal TUR for classes of systems that admit an upper bound on the asymmetry index.  Below, we demonstrate it for mesoscopic ballistic conductors, while in~\cite{SM} we derive a semiuniversal TUR~\footnote{
	Bounds for current rate functions for time-periodic Markovian systems are obtained in the very recent Ref.~\cite{Barato2018}. While the procedure there is general, the explicit bounds on current uncertainty are given only for time-independent contractions of elementary currents, and, thus, exclude basis currents of extracted work or heat. In this sense, our results are complementary to those in \cite{Barato2018}, and, moreover, provide the only known TUR to, for example, heat engines without time reversal~\cite{Risken1996,Schmiedl2008,Blickle2012,Martinez2015}, see discussion in \cite{SM}.
}
for periodically driven mesoscopic machines~\cite{Schmiedl2008,Blickle2012,Martinez2015,Brandner2015}.

Interestingly, for thermal machines with broken time-reversal symmetry, it is known that the diverging asymmetry index is necessary to achieve Carnot efficiency $\eta_C$ while maintaining finite power $P$ ~\cite{Brandner2013,Benenti2011,Brandner2013b}. On the other hand, the TUR~\eqref{TUR} has been recently related to the trade-off between power, efficiency, and constancy~\cite{Pietzonka2016c,Pietzonka2017b}, implying that $\eta_C$ for a time-reversible engine may be achieved at $P>0$ provided that fluctuations of power diverge, otherwise the power necessary vanishes, $P=0$. Our result~\eqref{result}, also allows for nonvanishing power when the asymmetry index diverges, see~\cite{SM}, consistently with~\cite{Brandner2013,Benenti2011,Brandner2013b}.    

Note that the breaking of the time-symmetric TUR~\eqref{max} by~\eqref{maxA} and~\eqref{maxF} is {\em not} a consequence of considering a particular linear combination of the basis currents.  Indeed, if we fix the coefficients $\c$, we can maximise the precision with respect to\ a choice of affinities [rather than a choice of coefficients as in \eqref{maxA}]. This optimal affinity is 
\begin{equation}\label{optF_LR}
\F_\text{opt} \, \propto\, \L_S^{-1}\L^\T \c= \c- \L_S^{-1}\L_A \c ,
\end{equation}
leading to a \emph{weaker} relation than~\eqref{max} 
 \begin{equation}\label{maxF_LR}
  \frac{J_\c^2}{\sigma D_\c} \leq
  \frac{\c^{\T}\L\,\L_S^{-1}\L^{\T}\c}
  {2\c^{\T}\L_S\c}=\frac{1}{2} +
  \frac{\c^{\T}\L_A\L_S^{-1}\L_A^\T\c}
  {2 \c^{\T}\L_S\c}
   .
\end{equation}

\begin{figure}
\includegraphics[width=\columnwidth]{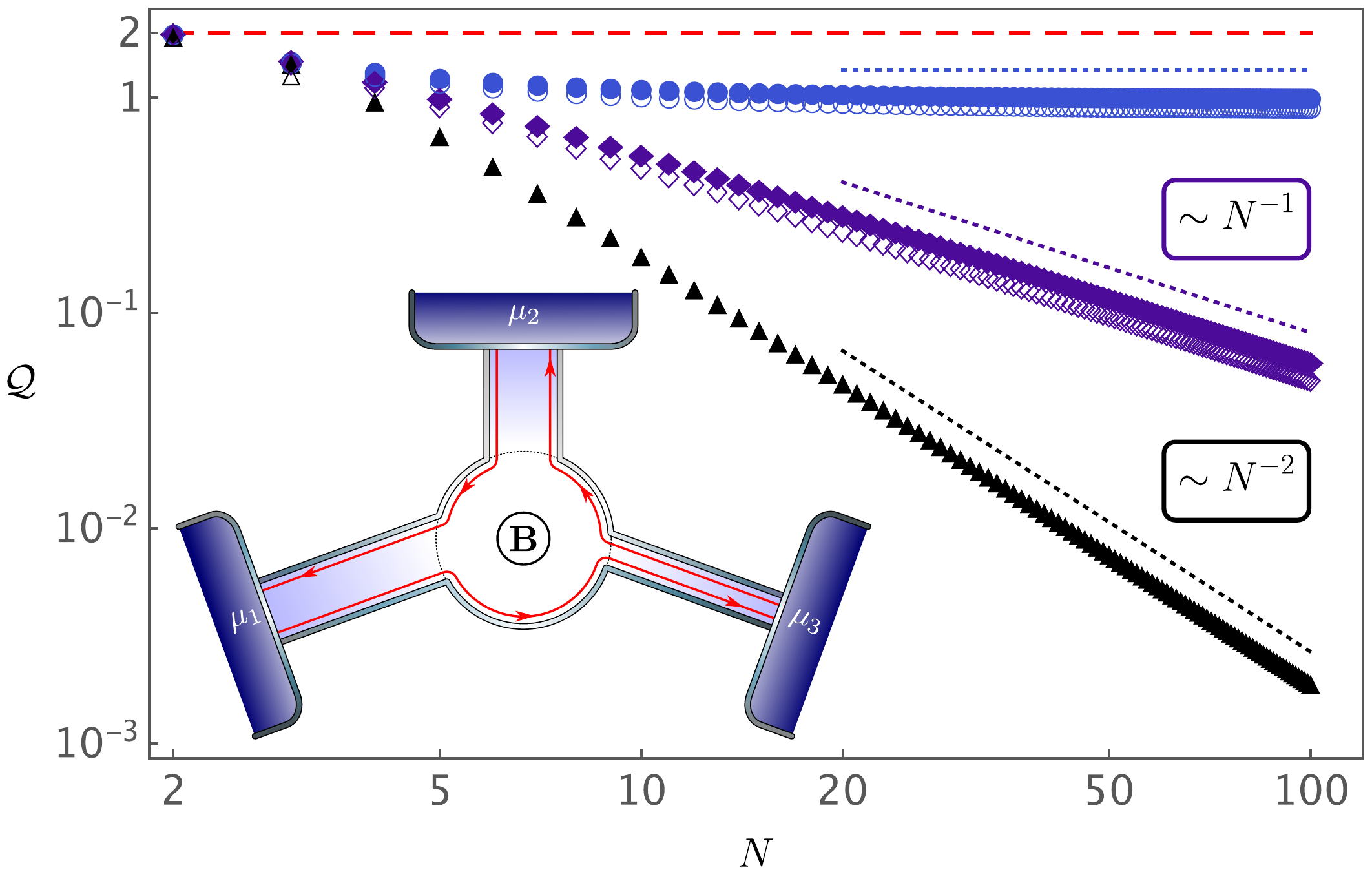}
\caption{Uncertainty products $Q$ for ballistic multiterminal transport as a function of $N$.
\emph{Inset}: Setup for $N=3$, with currents flowing along 
quantum Hall edge states (red lines).
\emph{Main figure:}  both $\Q_N$ for the most
precise basis current (blue circles: full --LR, empty --beyond), and $\Q_{\rm lin}$ for the
optimal current (purple diamonds: full --LR, empty --beyond) for linear bias profile break the time-reversible TUR~\eqref{TUR} (red dashed line). $\Q_{\rm sin}$ for sinusoidal bias (black triangles: full --LR, empty --beyond) saturates the LR bound \eqref{MC_TUR}. 
\label{PlotMC}}
\vspace*{-3mm}
\end{figure}

{\em Example.}---
As an application of our theory, we now discuss the ballistic 
transport of matter in mesoscopic multiterminal conductors. 
Such devices consist of a central junction connected to $N$
thermochemical reservoirs with common temperature $T$ and chemical
potentials $\mu_\alpha$ with $\alpha=1,\dots,N$, see 
Fig.~\ref{PlotMC}. 
For nonuniform affinities $F_\alpha\equiv(\mu_\alpha-\mu)/T$, where
$\mu$ is a reference chemical potential, the system is driven into a 
nonequilibrium steady state with finite particle currents $J_\alpha$
flowing in the individual terminals towards the junction.
The Onsager coefficients encoding the LR properties of the conductor 
can be obtained from the Landauer-B\"uttiker formula, 
$L_{\alpha\beta} =\int_0^\infty dE (\delta_{\alpha\beta}
-\mathcal{T}^{\alpha\beta}_{E\mathbf{B}})f_E$, which describes 
transport as the coherent quantum scattering of noninteracting 
particles \cite{Sivan1986,Baranger1989,Butcher1990,Buttiker1992,Matthews2014}.  
The energy-dependent transmission coefficients 
$0\leq\mathcal{T}^{\alpha\beta}_{E\mathbf{B}}\leq 1$ thereby contain
the scattering amplitudes connecting incoming and outgoing 
single-particle waves and $f_E\equiv \bigl( 2\cosh [(E-\mu)/(2T)]
\bigr)^{-2}$ denotes the derivative of the Fermi function. 
Here, the Planck and Boltzmann constants were set to $1$.

\smallskip 

For charged particles, the time-reversal symmetry of single-particle
scattering processes can be broken through an external magnetic field
$\B$.
The transmission coefficients, and, hence, the Onsager coefficients, 
are then typically not symmetric. 
However, the asymmetry index \eqref{sL} of the Onsager matrix is still
subject to the constraint~\cite{Brandner2013} 
\begin{equation}
\label{MC_BndAI}
s_{{{\rm MJ}}}\leq \cot(\pi/N),
\end{equation}
which follows from current conservation and gauge invariance requiring
the sum rules 
$\sum_\alpha \mathcal{T}^{\alpha\beta}_{E\mathbf{B}}=
\sum_\beta \mathcal{T}^{\alpha\beta}_{E\mathbf{B}}$
\cite{Blanter2000}.  
Our general result \eqref{result}, thus, implies the lower bound
\begin{equation}\label{MC_TUR}
\Q_{\c}\equiv \sigma D_\c/(J_\c)^2\geq 2\sin^2(\pi/N),
\end{equation}
on the product of the squared relative uncertainty of any current and
the rate of entropy production. We emphasize that the bound~\eqref{MC_TUR}, independent from the potential landscape inside the junction and the strength of an external magnetic field,  is \emph{valid for any mesoscopic conductor} with $N$ terminals, cf.~\eqref{MC_BndAI} and~\cite{Brandner2013} .

In Fig.~\ref{PlotMC} we consider a perfectly chiral junction, which
can be realized through a strong magnetic field enforcing
quantum Hall edge states \cite{Sothmann2014a,Sanchez2015a,
Sanchez2015b}. Assuming that only one edge state contributes to the transport 
process, the corresponding transmission coefficients are given by 
$\mathcal{T}^{\alpha\beta}_{E\mathbf{B}}=\delta_{\alpha(\beta-1)}$ and
the Onsager coefficients read $L_{\alpha\beta}=
\tau[\delta_{\alpha\beta}-\delta_{\alpha(\beta-1)}]$, where 
$\tau\equiv T/[1+\exp(-\mu/T)]$ \cite{Buttiker1992}, which corresponds to the maximal asymmetry
index~\eqref{MC_BndAI}. 

\smallskip
{\em (a) Linear bias}: First, we consider a linear bias landscape, 
i.e., $(\F_{{{\rm lin}}})_\alpha \equiv \FN \alpha/N$, where $\FN$ is 
an arbitrary constant. 
This choice leads to the uncertainty products $\Q_{\alpha<N}=N(N-1)$
and $\Q_N= N/(N-1)$ for the basis currents, which are bounded by $1$
rather than $2$, see Fig.~\ref{PlotMC} and \cite{SM}. This is due to the linear profile $\F_{{{\rm lin}}}$ being optimal,~\eqref{optF_LR}, for $N$th basis current, cf.~\cite{Brandner2017}. However, by combining the basis currents with the optimal 
coefficients for the linear profile,
$(\c_{{{\rm opt}}})_\alpha= \cN\bigl\{\alpha + [\alpha- (N+1)/2]^2 +
(1-N^2)/12\bigr\}$, which follow from \eqref{optC_LR} with 
$\cN \sim N^{-5/2}$ being the normalization factor, 
we obtain  $J_{{{\rm opt}}}=\tau \cN \FN (N^2-1)/6$ and 
$D_{{{\rm opt}}} =\tau \cNsq N(N^2-1)/3$~\cite{SM}. 
Hence, the minimal uncertainty product
$\Q_{{{\rm lin}}}\equiv\sigma D_{{{\rm opt}}}/(J_{{{\rm opt}}})^2
=6/(N+1)$, \emph{vanishes} for large $N$, see Fig.~\ref{PlotMC}. 
Notably, due to current conservation, both $\Q_{N}$ and 
$\Q_{{{\rm lin}}}$ saturate the general bound \eqref{MC_TUR} for 
the simplest case $N=2$, where the Onsager matrix is symmetric and
\eqref{TUR} holds, and for the minimal nonsymmetric case $N=3$ 
\cite{SM}.

\smallskip
{ \em (b) Optimal bias}: 
To saturate the bound \eqref{MC_TUR}, the bias profile also has 
to be optimized, cf.~\eqref{maxA} and~\eqref{maxF}. 
This procedure leads to the optimal affinities
$(\F_{{{\rm opt}}})_\alpha = \FF \cos(2\pi\alpha/N)$ with the 
corresponding rate of entropy production $\sigma=\FF^2\tau N 
\sin^2(\pi/N)$ \cite{SM}. 
For this bias landscape, the uncertainty products of the basis 
currents increase with the number of terminals \cite{SM}. 
However, for the optimal current given by \eqref{optC_LR} as $(\c_{{{\rm opt}}})_\alpha =\cN  \bigl[\cos(2\pi\alpha/N)+\cot(\pi/N)\sin(2\pi\alpha/N)]$, where $\cN\sim N^{-1}$ is the normalization factor, we have 
${J}_{{{\rm opt}}}= \tau\FF\,\cN N$
and ${D}_{{{\rm opt}}}=
2\tau\,\cNsq N$ \cite{SM}. 
Thus, the minimal uncertainty product ${\Q}_{{{\rm sin}}}$ saturates 
the bound \eqref{MC_TUR} and tends to zero as $N^{-2}$, 
see Fig.~\ref{PlotMC}. 
We note that, for $N=3$, $\Q_\text{lin}=\Q_\text{sin}$, since current 
conservation implies the equivalence of the linear and the sinusoidal 
bias landscape. 

\smallskip
{\em Variational principle and TUR beyond linear response}.--- 
The bound \eqref{maxA} can be extended beyond LR
using a variational principle for the relative uncertainty. 
To this end, first, we note that $J_\c^2/ D_\c=\max_x 
\left(-x^2  D_\c  + 2 x J_\c \right)$, where the rhs\ attains its 
maximum at $x=J_\c/D_\c$.
If we further maximize over $\c$ we get the optimal current among 
linear combinations of basis currents. 
Replacing $x \c$ with $\c$, we obtain  
\begin{equation}\label{varC}
\max_{\c} J_\c^2/ D_\c = 
\max_{\c} \left( -\, \c^\T\! \D \c\,   + 2 \c^{\T}\! \J \right).
\end{equation}
Here, $ (\mathbb{D})_{\alpha\beta}=D_{\alpha\beta}$ is the matrix of
correlations between the basis currents, and 
$(\mathbf{J})_\alpha=J_\alpha$ the vector of average currents, which 
is, in general, a nonlinear functions of $\F$. Moreover, in LR an analogous variational principle can be obtained for the optimal choice of affinities maximizing the precision of a given current in~\eqref{maxF_LR}~\cite{SM}.
By differentiating \eqref{varC} with respect to\ $\c$, we obtain the condition
$ \D\, \c_\text{opt}= \J$ on the optimal coefficients
$\c_{{{\rm opt}}}$. 
The relative uncertainty, $J_\c^2/ D_\c$, is invariant to multiplying $\c$ by a scalar, so the optimality
condition relaxes to
\begin{equation}\label{optC}
\D\, \c_\text{opt} \propto \J. 
\end{equation}
If $\D$ is invertible, \eqref{optC} leads to
$\c_\text{opt}\propto\D^{-1}\J $. 
In LR, this relation reduces to the condition \eqref{optC_LR} for
saturation of~\eqref{maxA}.
In general, the solution of \eqref{optC} exists only if $\J$ is
orthogonal to the kernel of $\D$; otherwise the maximum of \eqref{varC}
is infinite and the relative uncertainty is trivially bounded from 
below by zero, cf. \eqref{result} 
\footnote{In LR, the solution of~\eqref{optC} exists,
for arbitrary affinities $\F$, only if the kernel of $\L_S$ lies in 
the orthogonal complement of the range of $\L_A$. 
Note that, if the kernel of $\L_S$ overlaps with the range of $\L_A$,
the asymmetry index of $\L$ is infinite and \eqref{maxF} still holds
formally, see~\cite{SM} for details.}. 

In the former case, \eqref{optC} implies the identity
\begin{equation}\label{TURoptC}
\frac{1}{ \Q_\text{opt} }\equiv\max_{\c}\frac{J_\c^2}{\sigma D_\c}
= \frac{\J^\T  \D^{+} \J}{\F^\T\J} ,
\end{equation}
where $(\cdot)^+$ indicates the pseudoinverse. This relation~\eqref{TURoptC} can be further formally connected  to the asymmetry index in analogy to Eqs.~\eqref{maxF} and~\eqref{sL}, see \cite{SM} and~\cite{Vroylandt2018}.

\smallskip
{\em (i) Time-reversible case}: To the first-order beyond LR we have $\J=\L\F +\delta \J+\mathcal{O}(\F^2)$ and $\D= 2\L +\delta \D+\mathcal{O}(\F^2)$, so from~\eqref{TURoptC}
\begin{equation} 
\frac{J_\c^2}{\sigma D_\c}
\leq \frac{1}{2} 
 + \frac{2 \F^\T \delta \J-\F^\T\delta \D\F  }{4
	\F^\T \L \F} +\mathcal{O}(\F^2). \label{TUR1order}
\end{equation}
Both for homogeneous Markovian dynamics, and {\em for periodically
	driven Markovian systems with time-reversible protocols}, the first
correction in \eqref{TUR1order} vanishes, as $\delta \J=\delta \D\F/2$ due to
Gallavotti-Cohen symmetries \cite{Andrieux2007,Gaspard2013a,Ray2017}. The TUR in Eq.~\eqref{max}, thus, holds up to $\mathcal{O}(\F^2)$ for all $\F$ 
(except $\F$ in the kernel of $\L_S$).
Moreover, the entropy production rate remains the optimal current,  $\c_\text{opt}\propto\D^+\J=\F/2+\mathcal{O}(\F^2)$, with $\mathcal{Q}_\text{opt}=1/2+\mathcal{O}(\F^2) $. 
We note that the TUR in Eq.~\eqref{TUR} was derived beyond LR as a 
consequence of a quadratic bound on that rate function that also obeys
the Gallavotti-Cohen symmetry \cite{Gingrich2016,Gingrich2016b,
	Nardini2018}.
	
\smallskip
{\em (ii) Time-nonreversible case: example revisited}:
To explore Eq.~\eqref{TURoptC} without
time-reversal symmetry, we consider a chiral multiterminal junction
in the nonlinear regime. 
For simplicity, we focus on the semiclassical limit, where the 
density of carriers in the conductor is low such that Pauli blocking
and quantum correlations can be neglected \cite{Callen1985}. 
Under this condition, the mean currents and fluctuations can be derived as
$J_\alpha=\bar{\tau}(e^{F_\alpha}-e^{F_{\alpha+1}})$ and 
$D_{\alpha\beta}=
\bar{\tau}\delta_{\alpha\beta}(e^{F_\alpha}+e^{F_{\alpha+1}})
-\bar{\tau}\delta_{\alpha(\beta-1)}e^{F_\beta}
-\bar{\tau}\delta_{\beta(\alpha-1)}e^{F_\alpha}$, respectively,
where $\bar{\tau}\equiv T\exp[\mu/T]$ \cite{SM}. 
In Fig. \ref{PlotMC}, we show how the uncertainty product 
$\Q_\text{opt}$ for the optimal current given by \eqref{optC}
scales with $N$ for linear and sinusoidal bias profiles. 
For the linear profile, $(\F_{{{\rm lin}}})_\alpha \equiv\FF\alpha/N$,
choosing the amplitude $\FF$ to minimize $\Q_\text{opt}$ leads to 
$\Q_{\rm lin} \geq \psi^* 6/(N+1)$, with an additional factor $\psi^*\simeq 0.83$ compared to
LR, as occurs for the basis currents~\cite{Brandner2017}; see, also,~\cite{SM}. In contrast, for $N\geq 4$ and the sinusoidal bias profile
$(\F_{{{\rm sin}}})_\alpha\equiv\FF_1\cos(2\pi\alpha/N) 
+\FF_2\sin(2\pi\alpha/N)$,
the optimal amplitudes $\FF_1$ and $\FF_2$ are within the LR regime and the 
bound \eqref{MC_TUR} holds; see~\cite{SM} for details. As the sinusoidal bias profile is no longer guaranteed to be optimal beyond LR, only a systematic optimization of the bias profile would
lead to a general TUR for ballistic conductors beyond LR, which constitutes 
an interesting problem for future work.

\bigskip
We thank P. Pietzonka for comments on the first version of this work.
K.B. acknowledges financial support from the Academy of Finland (Contract No. 296073) and is affiliated with the Centre of Quantum Engineering. 
This work was supported by EPSRC Grant No. EP/M014266/1 (J.P.G., K.M.).

\bibliography{bounds_LR}

\section*{Supplemental Material}

\section{SM1: Mesoscopic Conductor}
\subsection{A. Linear Response} \label{sec:MJinLR}

We derive the expressions discussed in the main text for the 
uncertainty products of basis and optimal currents in a chiral 
conductor with linear and optimal bias landscape, respectively. 
To this end, first, we recall the general expression 
\begin{equation}\label{SM_LB}
L_{\alpha\beta} = \int_0^\infty \!\!\! dE \; 
\bigl(\delta_{\alpha\beta}-\TC_{E\B}^{\alpha\beta}\bigr)f_E
\end{equation}
for the Onsager coefficients describing a single-channel mesoscopic
conductor with multiple terminals in the ballistic regime 
\cite{Sivan1986,Baranger1989,Butcher1990,Buttiker1992}. 
The $\TC_{E\B}^{\alpha\beta}$ are thereby the energy and magnetic 
field dependent transmission coefficients of the junction and 
\begin{equation}
f_E\equiv 1/\bigl[2\cosh\bigl((E-\mu)/(2T)\bigr)\bigr]^2
\end{equation}
denotes the derivative of the Fermi function evaluated at the common
temperature $T$ of the reservoirs and the reference chemical potential
$\mu$. 
Note that, throughout this article, we consider spinless, 
noninteracting Fermions as transport carriers and Boltzmann's and
Planck's constant were set to 1. 
For later reference, we note that the Onsager matrix $\mathbb{L}$ 
defined through \eqref{SM_LB} fulfills 
\begin{equation}\label{SM_MCCons}
\mathbb{L}^T\mathbf{1}=0 \quad\text{and}\quad
\mathbb{L}\mathbf{1} = 0 
\quad\text{with}\quad\mathbf{1}\equiv (1,\dots,1)^T.
\end{equation}
These two relations respectively reflect current conservation and 
gauge invariance. 
On the microscopic level, they correspond to the sum rules 
\begin{equation}\label{MC_SR}
\sum_\alpha\TC^{\alpha\beta}_{E\mathbf{B}}= 1 
\quad\text{and}\quad
\sum_\beta \TC^{\alpha\beta}_{E\mathbf{B}}= 1,
\end{equation}
which the transmission coefficients $\TC^{\alpha\beta}_{E\mathbf{B}}$
have to obey for any fixed energy $E$ and magnetic field $\B$
\cite{Blanter2000}. 

In presence of a strong magnetic field, charged particles traverse 
a perfect mesoscopic conductor along chiral quantum Hall edge states
\cite{Buttiker1992}. 
For a single transport channel, the transmission coefficients 
describing this transport mechanism are given by 
\begin{equation}\label{SM_TC}
\mathcal{T}^{\alpha\beta}_{E\mathbf{B}} = \delta_{\alpha(\beta-1)},
\end{equation}
where $\alpha,\beta$ are periodic indices and the direction of the
magnetic field has been chosen such that the edge states circulate
counterclockwise, see Fig.~1 of the main text. 
Inserting \eqref{SM_TC} into \eqref{SM_LB} yields
\begin{equation}\label{MC_ChiraOC}
L_{\alpha\beta}=
\tau\left[\delta_{\alpha\beta}-\delta_{\alpha(\beta-1)}\right]
\end{equation}
with $\tau\equiv T/[1+\exp(-\mu/T)]$. 

{\bf 1. Linear bias.}
For a linear bias landscape, i.e., 
\begin{equation}\label{MC_LinBias}
(\F_{{{\rm lin}}})_\alpha = \FF \alpha/N
\end{equation}
with $\FF$ being an arbitrary constant,
\eqref{MC_ChiraOC} leads to the expressions 
\begin{align}
\label{SM_MCBasis}
& J_{\alpha<N} = -\tau\FF/N, \; 
J_N = \tau \FF(1-1/N)  \;\;\text{and}\\
& D_\alpha = 2\tau \nonumber
\end{align}
for the basis currents and the corresponding fluctuations, 
respectively. 
The rate of entropy production becomes 
\begin{equation}\label{SM_MCEnt}
\sigma = \tau \FF^2 (1-1/N)/2
\end{equation}
and the uncertainty products for the basis currents are thus given
by 
\begin{equation}\label{SM_MCUPB}
\Q_{\alpha<N} = N(N-1) \quad\text{and}\quad
\Q_{N} = N/(N-1). 
\end{equation}

The optimal current minimizing the uncertainty product is obtained by
contracting the basis currents \eqref{SM_MCBasis} with the 
coefficients 
\begin{equation}\label{SM_OptCoeff}
\c_{{{\rm opt}}} \propto \mathbb{L}_S^{-1}\mathbb{L}\F
= \F + \mathbb{L}_S^{-1}\mathbb{L}_A\F,
\end{equation}
see Eq. (7) of the main text. 
Inserting \eqref{MC_ChiraOC} into this expression yields 
\begin{equation}\label{MC_LinBOptC}
(\c_{{{\rm opt}}})_\alpha= 
\cN\bigl[\alpha + (\alpha- (N+1)/2)^2 + (1-N^2)/12\bigr],
\end{equation}
where a normalization factor $\cN\sim N^{-5/2}$ has been introduced
such that $\c_{{{\rm opt}}}^\T\c_{{{\rm opt}}}=1$. 
We note that, as a consequence of \eqref{SM_MCCons}, neither
the currents $\J$ nor the optimal contraction coefficients 
$\c_{{{\rm opt}}}$ are affected by a uniform shift of the bias 
landscape. 

From \eqref{MC_LinBOptC} and \eqref{SM_MCBasis}, we obtain the optimal
current and its fluctuations as 
\begin{align}
J_{{{\rm opt}}} &=\tau \cN \FF (N^2-1)/6 \quad\text{and}\\ 
D_{{{\rm opt}}} &=\tau \cNsq N(N^2-1)/3.\nonumber
\end{align}
Using \eqref{SM_MCEnt}, we arrive at the minimal uncertainty
product for a chiral conductor with $N$ terminals driven by a linear
bias landscape, 
\begin{equation}\label{MC_ChiraUPMin}
\Q_{{{\rm lin}}}\
\equiv\sigma D_{{{\rm opt}}}/(J_{{{\rm opt}}})^2
=6/(N+1). 
\end{equation}
In contrast to the uncertainty products \eqref{SM_MCUPB} for the basis
currents, the minimized uncertainty product \eqref{MC_ChiraUPMin} 
decays to zero as the number of terminals $N$ increases, see Fig.~1 of
the main text. 
Finally, we note that, for $N=2$ and $N=3$, we have 
$\Q_{{{\rm lin}}}=Q_N=N/(N-1)$; in both cases the vector 
$\c_{{{\rm opt}}}$ is proportional to the $N^{{{\rm th}}}$ basis 
vector up to a uniform shift, which does not alter the uncertainty 
product due to the properties \eqref{SM_MCCons} of the Onsager matrix.

{\bf 2. Optimal bias.} 
An optimal bias profile $\F_{{{\rm opt}}}$, for which the bound Eq. 
(2) of the main text can be saturated, is generally given by a linear
combination of extremal eigenvectors of the Hermitian matrix
\begin{equation}
\X = \mathbb{L}^{-1/2}_S(i\mathbb{L}_A)\mathbb{L}_S^{-1/2},
\end{equation}
that is, those eigenvectors corresponding to either the minimal or
maximal eigenvalue of $\X$. 
To find $\F_{{{\rm opt}}}$ for a chiral conductor, first, we note that,
due to the rotational symmetry of the system, the symmetric and 
antisymmetric part of the Onsager matrix defined through
\eqref{MC_ChiraOC} share a common set of normalized eigenvectors 
$\v_\kappa$ with components 
\begin{equation}\label{MC_ChiraEVec}
(\v_\kappa)_\alpha = 
\exp\bigl(i [2\pi \kappa/N]\, \alpha\bigr)/\sqrt{N}.
\end{equation}
Here, $\kappa=1,\dots,N$ and the corresponding eigenvalues of 
$\mathbb{L}_S$ and $\mathbb{L}_A$ read 
\begin{equation}\label{MC_ChiraEVal}
\lambda^S_\kappa = 2\sin^2(\pi\kappa/N) 
\quad\text{and}\quad
\lambda^A_\kappa = -i\sin(2\pi\kappa/N),
\end{equation}
respectively. 
This result shows that, although $\mathbb{L}$ is not invertible, the
bounds Eq. (6), Eq. (8) and hence Eq. (9) of the main text still hold
if $\mathbb{L}^{-1}$ is understood as a pseudoinverse throughout, 
since the kernel of $\mathbb{L}_S$ is orthogonal to the range of 
$\mathbb{L}_A$, see Sec.~\ref{sec:var}; 
the joint null-space of $\mathbb{L}_S$ and $\mathbb{L}_A$ is spanned by
the uniform vector $\mathbf{1}=\sqrt{N}\v_0$. 
Note that, first, this relation was exploited implicitly already when
we used \eqref{SM_OptCoeff} to obtain \eqref{MC_LinBOptC}.
Second, the same argument applies in general to any Onsager matrix 
that has a one-dimensional null-space and obeys \eqref{SM_MCCons}. 

Using \eqref{MC_ChiraEVec} and \eqref{MC_ChiraEVal} and noting that 
the $\v_{\kappa>0}$ are also the eigenvectors of the matrix $\X$ with 
corresponding eigenvalues $i\lambda^A_\kappa/\lambda^S_\kappa$, we 
obtain the general form of the optimal bias profile for a chiral
conductor,
\begin{align}\label{MC_ChiraOptF}
(\F_{{{\rm opt}}})_\alpha &= 
\FF_1 \cos(2\pi\alpha/N) +\FF_{2} \sin(2\pi\alpha/N)\nonumber\\
&\equiv \FF \sin(2\pi\alpha/N+\phi),
\end{align}
where $\FF_1$ and $\FF_2$ are arbitrary constants. 
Note that, in the main text, we have set $\phi=\pi/2$ for simplicity. 
The basis currents for the optimal bias landscape \eqref{MC_ChiraOptF}
are 
\begin{equation}
J_\alpha = -2\tau\FF\cos\bigl(\pi(1 + 2\alpha)/N + \phi\bigr)
                      \sin\bigl(\pi/N\bigr)
\end{equation}
with fluctuations $D_\alpha=2\tau$ and the corresponding rate of 
entropy production,
\begin{equation}\label{MC_ChiraOptFEnt}
\sigma = \tau\FF^2 N \sin^2\bigl(\pi/N\bigr),
\end{equation}
does not depend on the parameter $\phi$. 
Hence, the uncertainty products, 
\begin{equation}\label{MC_QUPSin}
\mathcal{Q}_\alpha =
N/\bigl[2\cos^2\bigl(\pi(1 + 2\alpha)/N + \phi\bigr)\bigr],
\end{equation}
of the basis currents diverge at least linearly in the number of
terminals $N$. 

Finally, using the condition \eqref{SM_OptCoeff}, we obtain the 
optimal contraction coefficients,
\begin{align}\label{MC_ChiraOptFC}
&(\c_{{{\rm opt}}})_\alpha =\cN  \bigl[\sin(2\pi\alpha/N+\phi)\\
&\hspace*{2.5cm}-\cot(\pi/N)\cos(2\pi\alpha/N+\phi)],\nonumber
\end{align}
for a chiral conductor driven by the sinusoidal bias landscape 
\eqref{MC_ChiraOptF}, where the normalization factor $\cN\sim
N^{-1}$ is chosen such that $\c_{{{\rm opt}}}^T\c_{{{\rm opt}}}=1$.
The corresponding optimal current and fluctuations, 
\begin{equation}
J_{{{\rm opt}}} = \tau \FF \cN  N
\quad\text{and}\quad
D_{{{\rm opt}}} = 2\tau \cN^2 N,
\end{equation}
are independent of $\phi$ and their uncertainty product 
\begin{equation}\label{MC_ChiraOptQ}
\mathcal{Q}_{{{\rm sin}}}= 2\sin^2(\pi/N)
\end{equation} 
saturates the bound Eq. (12) of the main text. 
Furthermore, in contrast to the uncertainty products of the basis 
currents \eqref{MC_QUPSin}, $\mathcal{Q}_{{{\rm sin}}}$ decays 
monotonically to $0$ as $N$ increases. 
We conclude this section by noting that, for $N=2$, we have 
$\F_{{{\rm lin}}} \propto \F_{{{\rm opt}}}$ up to an irrelevant
homogeneous shift. 
The same equivalence holds for $N=3$. 
This observation explains why $\mathcal{Q}_{{{\rm sin}}} =
\mathcal{Q}_{{{\rm lin}}}$ for $N=2,3$, see Fig.~1 of the main text.

\subsection{B. Nonlinear Regime} \label{sec:MJbeyondLR}

\begin{figure*}
\includegraphics[width=2\columnwidth]{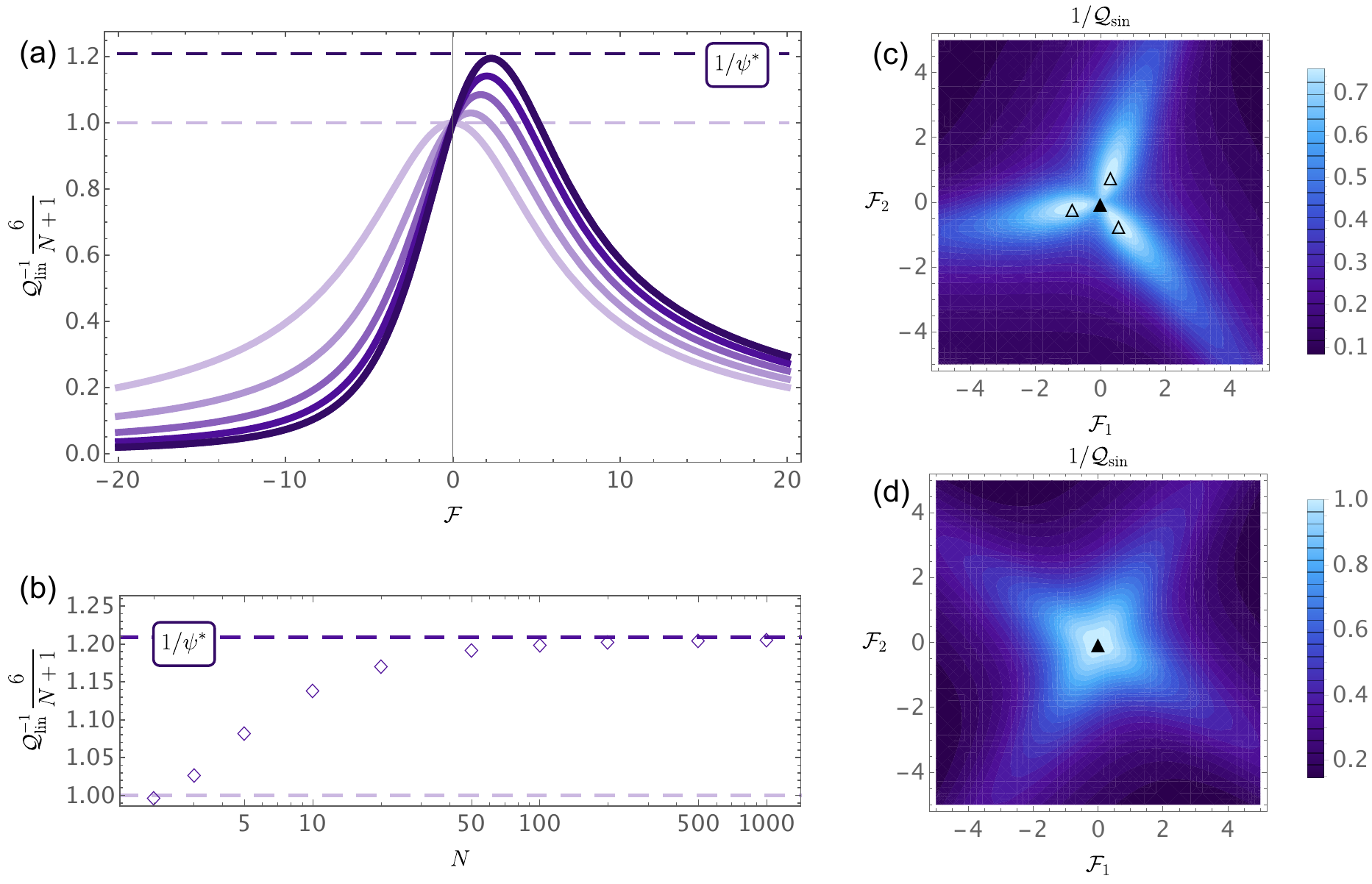}
\caption{Uncertainty products of the optimal current for chiral 
conductors with different bias landscapes and numbers of terminals. 
{\bf Linear bias:} (a) Plot of $1/\Q_{\rm lin}$ as a function of the 
strength $\FF$ of the bias profile \eqref{MC_LinBias} for the 
symmetric case $N=2$ and  $N=3,5,10,100,1000$, where darker color 
corresponds to larger $N$.
The dashed lines indicate the scaling factors within (light blue) and
beyond (dark blue) linear response. 
(b) Scaling of the maximum of $1/\Q_{\rm lin}$ with respect to$\FF$
as a function of the number of terminals $N$. 
{\bf Sinusoidal bias:} Plot of $1/\Q_{\rm sin}$ as a function of 
the amplitudes $\FF_1$ and $\FF_2$ in the bias profile 
\eqref{MC_ChiraOptF} for $N=3$ in (c) and for $N=4$ in (d).
Both plots show an $N$-fold rotational symmetry, which corresponds to
the symmetry of the underlying system.
For $N=3$, the uncertainty product $\Q_{\rm sin}$ becomes minimal 
when $\FF_1$ and $\FF_2$ are of order 1 (black empty triangles),
while for $N=4$ the minimum is attained within linear response
(black triangles), see, also, Fig.~1 of the main text. 
\label{PlotSM}}
\vspace*{-3mm}
\end{figure*}

Beyond linear response, scattering theory provides the general
expression \cite{Buttiker1992}
\begin{equation}
J_\alpha = \int_0^\infty \!\!\! dE 
           \sum_\beta \TC_{E\B}^{\alpha\beta}
           \bigl[\f^\alpha_E - \f^\beta_E\bigr]
\end{equation}
for the steady-state currents flowing in a ballistic multiterminal
conductor, where 
\begin{equation}
\f^\alpha_E = 1/\bigl[1+\exp\bigl([E-\mu_\alpha]/T\bigr)\bigr]
\end{equation}
is the Fermi distribution of the reservoir $\alpha$.
The corresponding fluctuations,
\begin{equation}
D_{\alpha\beta} = D_{\alpha\beta}^{{{\rm cl}}} 
                 -D_{\alpha\beta}^{{{\rm qu}}},
\end{equation}
can be divided into a quasi-classical part 
$D_{\alpha\beta}^{{{\rm cl}}}$ and a quantum part 
$D_{\alpha\beta}^{{{\rm qu}}}$. 
For a single-channel conductor, the explicit expressions for these 
two contributions read \cite{Buttiker1992}
\begin{multline}
D_{\alpha\beta}^{{{\rm cl}}} \equiv
\delta_{\alpha\beta} \int_0^\infty \!\!\! dE
\sum_\gamma \TC^{\alpha\gamma}_{E\B}
\bigl[ \f^\alpha_E(1-\f^\gamma_E)+\f^\gamma_E(1-\f^\alpha_E)\bigr]\\
-\int_0^\infty\!\!\! dE \bigl[
\TC^{\alpha\beta}_{E\B} \;  \f^\beta_E(1-\f^\beta_E)
+\TC^{\beta\alpha}_{E\B} \; \f^\alpha_E(1-\f^\alpha_E) \bigr]
\end{multline}
and 
\begin{multline}
D^{{{\rm qu}}}_{\alpha\beta}
\equiv \int_0^\infty\!\!\! dE \sum_{\gamma\delta}
(S^{\alpha\gamma}_{E\B})^\ast (S^{\beta\delta}_{E\B})^\ast
S^{\alpha\delta}_{E\B} S^{\beta\gamma}_{E\B}\\
\times
\bigl[\f^\alpha_E-\f^\gamma_E\bigr]\bigl[\f^\beta_E-\f^\delta_E\bigr],
\end{multline}
where $S^{\alpha\beta}_{E\B}$ is the scattering amplitude connecting
an incoming wave in the terminal $\beta$ to an outgoing one in the 
terminal $\alpha$ at fixed energy $E$ and magnetic field $\B$. 
These objects are directly related to the transmission coefficients
$\TC^{\alpha\beta}_{E\B} = \bigl|S^{\alpha\beta}_{E\B}\bigr|^2$.
Note that only the quasi-classical fluctuations
$D^{{{\rm cl}}}_{\alpha\beta}$ can be expressed solely in terms of 
transmission coefficients, while the quantum term
$D^{{{\rm qu}}}_{\alpha\beta}$, which describes correlations resulting
from the antisymmetry of multi-particle states, depends intrinsically
on the scattering amplitudes.  

For simplicity, in this article, we focus on the semiclassical 
regime, where the density of carriers in the conductor is small such
that Pauli blocking and exchange interactions become irrelevant
\cite{Callen1985}. 
Specifically, we assume that fugacities 
\begin{equation}
\varphi_\alpha \equiv \exp(\mu_\alpha/T)
\end{equation}
of the individual reservoir are much smaller than 1. 
Neglecting second-order corrections in $\varphi_\alpha$, we thus 
arrive at the simplified expressions 
\begin{equation}\label{MC_SClassJ}
J_\alpha = \int_0^\infty \!\!\! dE 
\sum_\beta \TC^{\alpha\beta}_{E\B}\bigl[u^\alpha_E -u^\beta_E\bigr]
+ \mathcal{O}\bigl(\varphi^2\bigr)
\end{equation}
and 
\begin{multline}\label{MC_SClassD}
D_{\alpha\beta} = 
\int_0^\infty \!\!\! dE \; u_E \biggl\{
\delta_{\alpha\beta}\sum_\gamma \TC^{\alpha\gamma}_{E\B}
\bigl[u^\alpha_E+u^\gamma_E\bigr]\\
-\TC^{\alpha\beta}_{E\B}u^\beta_E
-\TC^{\beta\alpha}_{E\B}u^\alpha_E
\biggr\} + \mathcal{O}\bigl(\varphi^2\bigr)
\end{multline}
for the mean currents and current fluctuations, respectively. 
Here, 
\begin{equation}
u^\alpha_E\equiv \exp\bigl(-[E-\mu_\alpha]/T\bigr)
\end{equation}
denotes the classical Maxwell-Boltzmann distribution. 
Note that \eqref{MC_SClassD} depends only on transmission 
coefficients and not on scattering amplitudes, because the quantum
fluctuation term $D^{{{\rm qu}}}_{\alpha\beta}$ vanishes in first 
order with respect tothe fugacities.

\textbf{Chiral conductor.}
Inserting the transmission coefficients \eqref{SM_TC} for a chiral
multiterminal conductor into \eqref{MC_SClassJ} and 
\eqref{MC_SClassD} yields 
\begin{equation}
J_\alpha = \bar{\tau}\bigl(e^{F_\alpha}-e^{F_{\alpha+1}}\bigr)
\end{equation}
and 
\begin{multline}\label{MC_SChiraD}
D_{\alpha\beta} = 
\delta_{\alpha\beta}
\bar{\tau}\bigl(e^{F_\alpha}-e^{F_{\alpha+1}}\bigr)\\
-\bar{\tau}\delta_{\alpha(\beta-1)}e^{F_\beta}
-\bar{\tau}\delta_{\beta(\alpha-1)}e^{F_\alpha},
\end{multline}
where $\bar{\tau}\equiv T\exp[\mu/T]$ and 
$F_\alpha \equiv (\F)_\alpha = (\mu_\alpha-\mu)/T$. 
From \eqref{MC_SChiraD}, the contraction coefficients for the optimal
current can be found using Eq. (14) of the main text.
In Fig.~\ref{PlotSM} (a), we show plots of the corresponding
uncertainty product $\mathcal{Q}_{{{\rm lin}}}$ for a linear bias 
profile $(\F_\text{lin})_\alpha \equiv \FF \alpha/N$ as a function
of the bias strength $\FF$. 
Choosing $\FF$ such that the uncertainty product becomes minimal,
we find that $\mathcal{Q}_{{{\rm lin}}}$ decays as $1/N$ for 
large numbers of terminals, where, compared to the linear-response
regime, an additional scaling factor $\psi^\ast$ appears,
see Fig.~\ref{PlotSM} (b). 

\begin{figure}[htb!]
\includegraphics[width=\columnwidth]{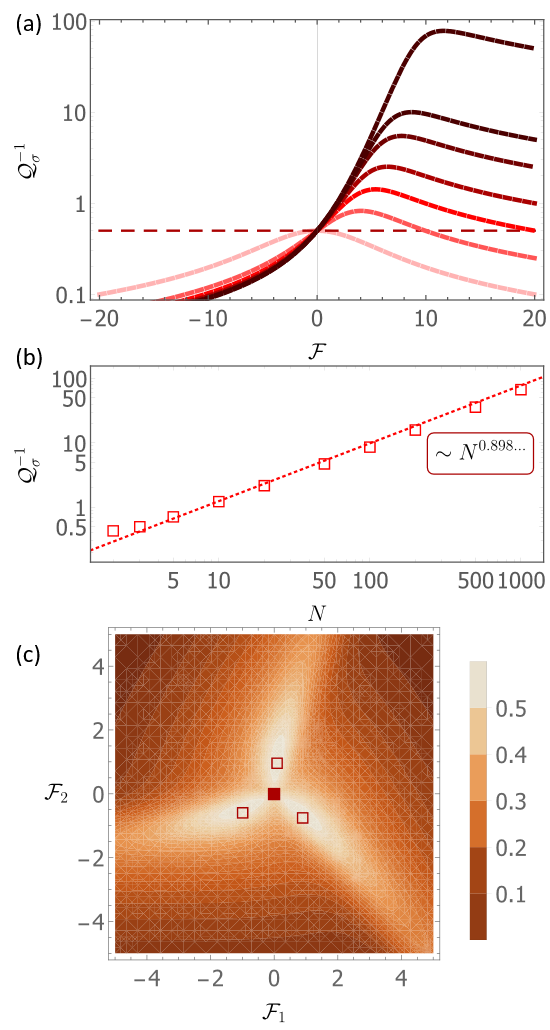}
\caption{Uncertainty product $\Q_\sigma$ of the rate of entropy 
production for chiral conductors with different bias landscapes and 
numbers of terminals. 
{\bf Linear bias:} (a) Plot of $1/\Q_\sigma$ as a function of the 
amplitude $\FF$ of the bias profile \eqref{MC_LinBias} for the 
symmetric case $N=2$ and  $N=3,5,10,20,50,100,1000$; darker color 
corresponds to larger $N$.
The dashed line indicates the linear response bound $1$. 
(b) Scaling of the maximum of $1/\Q_{\rm lin}$ with respect to$\FF$
as a function of the number of terminals $N$. 
{\bf Sinusoidal bias:} (c) Plot of $1/\Q_\sigma$ as a function of 
the amplitudes $\FF_1$ and $\FF_2$ in the bias profile 
\eqref{MC_ChiraOptF} for $N=3$. 
For $N\geq 4$ the minimum uncertainty product equals $1$ and is 
attained within linear response. 
\label{PlotSM2}}
\vspace*{-3mm}
\end{figure}

Using the sinusoidal bias profile \eqref{MC_ChiraOptF}, the
linear-response bound on the uncertainty product, Eq. (12) of the main
text, can be overcome  beyond linear response for $N=3$, 
see Fig.~\ref{PlotSM} (c).
However, for $N\geq 4$, $\mathcal{Q}_{{{\rm sin}}}$ still attains its
minimum at small bias amplitudes $\FF_1$ and $\FF_2$, for which 
the linear-response analysis applies, see Fig.~\ref{PlotSM} (d).  
This result suggest that, for chiral conductors with at least 4
terminals, Eq. (12) of the main text might still provide a lower
bound on the uncertainty product of the optimal current, even beyond
linear response. 
However, concluding this problem would require a systematic 
optimization of the bias profile in the nonlinear regime, which 
is beyond the scope of this work. 

\textbf{Role of entropy production.}
A second thermodynamic current of special interest, besides the 
optimal one, is given by the rate of entropy production
\begin{equation}
\sigma = \sum_\alpha F_\alpha  J_\alpha.
\end{equation}
In linear response, the corresponding uncertainty product $Q_\sigma=
\Delta^2_\sigma/\sigma$ is equal to $1/2$ irrespective of the specific
bias profile, see Eq.~(6).
Thus, in contrast to the optimal current, $\sigma$ generally obeys the
TUR for time-reversible systems, see Eq.~(4). 
Beyond linear response, however, this universal feature is lost as 
$\sigma$ is no longer independent from the number $N$ of terminals; 
for the linear bias profile, the uncertainty product $Q_\sigma$ can
even approach $0$ provided the bias strength $\FF$ is chosen 
suitably, see Fig.~\ref{PlotSM2}(a). 
The decay of $\mathcal{Q}_\sigma$ with increasing number of terminals
$N$ is, however, slower than the minimal uncertainty product
$\mathcal{Q}_{{{\rm lin}}}$, see Fig.~\ref{PlotSM2}(b). 
Hence, in this example, the rate of entropy production is not the 
thermodynamic current with minimal uncertainty product. 
Finally, we note that, for the sinusoidal profile, both 
$\mathcal{Q}_\sigma$ and the minimal uncertainty product
$\mathcal{Q}_{{{\rm sin}}}$ break the linear response bound, Eq.(4), 
only for $N=3$ as shown in Fig.~\ref{PlotSM2}(c) and Fig.~\ref{PlotSM}
(c,d), respectively.

\section{SM2: Periodic Thermodynamics in Linear Response}

\newcommand{\Tcal}{\mathcal{T}}
\newcommand{\ttauint}{\int_0^\Tcal\!\!\! dt
                      \int_0^\infty \!\!\! d\tau\;}
\newcommand{\Av}[1]{\bigl\langle #1\bigr\rangle}
\newcommand{\df}{\boldsymbol{\xi}}

Mesoscopic machines driven by periodic heating and mechanical 
forces form a second import class of systems that is covered by our 
theory.
In this section, we show that the asymmetry index of the corresponding
Onsager matrix is subject to a universal bound, which applies to both 
the classical and the quantum regime. 
As a key application, this result leads to constraints on the 
performance of mesoscopic heat engines that are significantly stronger
than the ones following from the basic laws of thermodynamics

\subsection{A. Setup}
In the classical realm, a generic mesoscopic device can be modeled as
a working system with degrees of freedom $\df$ and Hamiltonian 
\begin{equation}\label{PT_Cl_Hamiltonian}
H_t(\df) = H(\df) + T F_w f^w_t G^w(\df).
\end{equation}
Here, $H(\df)$ is the Hamiltonian of the free system, $f^w_t$ denotes
a dimensionless periodic control field coupling to the system variable
$G_w(\df)$.
The affinity $F_w$ determines the  amplitude of the external
perturbation. 
Thermal energy is provided to the system by a heat source periodically
increasing the temperature of its environment, 
\begin{equation}
T_t = T+ T^2 F_q f^q_t
\end{equation}
above the equilibrium value $T$, where $f^q_t$ is a dimensionless 
protocol and the second affinity $F_q$ controls the strength
of the temperature variation. 

Provided that the time evolution of the working system is governed by
overdamped Langevin dynamics, the kinetic coefficients connecting the
affinities $\FF_\alpha$ to the corresponding thermodynamic currents,
the work flux $J_w$ and the heat uptake $J_q$ \cite{Brandner2015}, 
are given by 
\begin{equation}\label{PT_Cl_OC}
L_{\alpha\beta} = \frac{1}{\Tcal}\ttauint \dot{f}^\alpha_t
                  f^\beta_{t-\tau} \dot{C}^{\alpha\beta}_\tau
                  \qquad (\alpha,\beta=w,q). 
\end{equation}
Here, $\Tcal$ denotes the length of one driving period and 
\begin{equation}\label{PT_Cl_CorrF}
C^{\alpha\beta}_t\equiv
 \bigl\langle \hat{G}^\alpha_t \hat{G}^\beta_0\bigr\rangle
-\bigl\langle \hat{G}^\alpha_t\bigr\rangle
 \bigl\langle \hat{G}^\beta_0 \bigr\rangle. 
\end{equation}
an equilibrium correlation function. 
The time-evolved system variables $\hat{G}^\alpha_t(\df)$ are thereby
determined through the equations of motion 
\begin{equation}\label{PT_Cl_EOM}
\partial_t \hat{G}^\alpha_t(\df) = \mathsf{L}(\df)\hat{G}^\alpha_t(\df)
\quad\text{with} \quad
\hat{G}^\alpha_0(\df) = G^\alpha(\df),
\end{equation}
where the free Fokker-Planck operator $\mathsf{L}(\df)$ describes the 
dynamics of the unperturbed system, $G^w(\df)$ was defined in
\eqref{PT_Cl_Hamiltonian} and $G^q(\df)\equiv -H(\df)/T$. 
Furthermore, the angular brackets in \eqref{PT_Cl_CorrF} indicate the
canonical average with respect tothe unperturbed Hamiltonian and
temperature, $H$ and $T$, for details see \cite{Brandner2015,
Risken1996}.

To further analyze the Onsager coefficients \eqref{PT_Cl_OC}, it is 
instructive to express them in terms of a mode expansion.  
To this end, first, we note that, assuming all degrees of freedom $\df$
are even under time reversal, the operator $\mathsf{L}(\df)$ satisfies
the detailed balance condition 
\begin{equation}
\bigl\langle (\mathsf{L} A)B\bigr\rangle
=\bigl\langle A (\mathsf{L}B)\bigr\rangle,
\end{equation}
where $A(\df)$ and $B(\df)$ are arbitrary system variables. 
This relation implies the existence of a complete set of eigenvectors
$V_\nu(\df)$ fulfilling
\begin{equation}
\mathsf{L}(\df)V_\nu(\df) = -\lambda_\nu V_\nu(\df) 
\quad\text{and}\quad
\bigl\langle V_\nu V_\mu\bigr\rangle = \delta_{\nu\mu}
\end{equation}
with $V_0(\df)=1$, $\lambda_0=0$ and $\lambda_{\nu\geq 1}>0$. 
Hence, from \eqref{PT_Cl_EOM} we obtain
\begin{equation}
\hat{G}^\alpha_t(\df) - \bigl\langle \hat{G}^\alpha_t\bigr\rangle
=\sum_{\nu\geq 1} d_\nu^\alpha V_\nu(\df)e^{-\lambda_\nu t}
\end{equation}
with expansion coefficients 
\begin{equation}
d_\nu^\alpha \equiv \Av{V_\nu G^\alpha}.
\end{equation}

After inserting this decomposition and the Fourier expansion 
\begin{equation}
f^\alpha_t = \sum_{n\in\mathbb{Z}} c^\alpha_n e^{in\Omega t},
\end{equation}
into \eqref{PT_Cl_CorrF} and \eqref{PT_Cl_OC}, we arrive at 
\begin{equation}\label{PT_Cl_Mode}
L_{\alpha\beta} = \sum_{\nu\geq 1}\sum_{n\in\mathbb{Z}}
d^\alpha_\nu d^\beta_\nu c_n^\alpha c_n^{\beta\ast} 
\frac{in \Omega \lambda_\nu}{in\Omega-\lambda_\nu},
\end{equation}
where $\Omega\equiv 2\pi/\calT$. 

To bound the asymmetry index of the Onsager coefficients 
\eqref{PT_Cl_OC}, we consider the quadratic form 
\begin{equation}\label{PT_Cl_AI1}
Q(s)\equiv \sum_{\alpha,\beta=w,q} 
\frac{s+i}{2}L_{\alpha\beta}  z_\alpha z_\beta^\ast
+\frac{s-i}{2}L_{\beta\alpha} z_\alpha z_\beta^\ast,
\end{equation}
where $s>0$ and $z_\alpha\in\mathbb{C}$. 
Inserting the mode expansion \eqref{PT_Cl_Mode} into 
\eqref{PT_Cl_AI1} yields 
\begin{equation}\label{PT_Cl_AI1}
Q(s) = \sum_{\nu\geq 1}\sum_{n\in\mathbb{Z}}
\frac{\lambda_\nu \Omega |Z^\nu_n|^2}{n^2\Omega^2+\lambda_\nu^2}
\bigl(s n^2\Omega + n\lambda_\nu\bigr),
\end{equation}
where
\begin{equation}
Z^\nu_n \equiv\sum_{\alpha=w,q} d^\alpha_\nu c^\alpha_n z_\alpha. 
\end{equation}
Since $\lambda_{\nu\geq 1}>0$, \eqref{PT_Cl_AI1} shows that the 
quadratic form $Q(s)$ is positive semidefinite for any 
$s\geq \lambda/\Omega$, where 
\begin{equation}
\lambda\equiv \max_{\nu \;|\; d^w_\nu \cdot d^q_\nu \neq 0} \lambda_\nu
\end{equation}
is the maximal relaxation rate of the free system that is sensitive to 
both perturbations $G^w$ and $G^q$;
the inverse rate $1/\lambda$ provides a lower bound on the typical 
decay time of the cross correlation $C^{wq}_t = C^{qw}_t$. 
Consequently, the asymmetry index of the Onsager coefficients 
\eqref{PT_Cl_OC} fulfills 
\begin{equation}\label{PT_Cl_BoundAI}
s_{{{\rm MM}}}\leq \lambda/\Omega. 
\end{equation}

This result entails that the uncertainty product of any thermodynamic
current accompanying the cyclic operation of the mesoscopic machine 
described by the coefficients \eqref{PT_Cl_OC} is subject to a lower 
bound proportional to $\Omega^2$. 
Thus, the faster the device is operated, the larger the 
uncertainty in its output must be. 
A perfectly precise machine, whose output does not fluctuate, can be 
realized only in the adiabatic limit $\Omega\rightarrow 0$, where the
Onsager coefficients \eqref{PT_Cl_OC} become antisymmetric, i.e., 
$L_{\alpha\beta}^{{{\rm ad}}} = - L_{\beta\alpha}^{{{\rm ad}}}$
\cite{Brandner2015}, and the bound \eqref{PT_Cl_BoundAI} diverges.

Very recently, a general set of bounds on the rate functions of currents in
time-periodic Markov systems was derived in Ref.\  \cite{Barato2018} using the level-2.5
formalism. 
For time-periodic dynamics close to equilibrium one can ask how the TURs obtained from this approach compare to the ones we have obtained here within LR. 
However, it is important to note that while the bounds on the rate functions of Ref.\ 
\cite{Barato2018} in principle can be applied to any current, 
the bounds on the current uncertainty, i.e., TURs,  follow, due the ansatz used for the minimisations in Ref.\ 
\cite{Barato2018}, only for currents that are time-independent
contractions of the elementary time-dependent currents in the system,
i.e., those given by the time-averaged fluxes between configurations.
In particular, these explicit bounds of Ref.\ \cite{Barato2018} {\em do not apply in
	general to basis currents} $J_\alpha$, i.e., to the fluxes conjugate to
the affinities, such as extracted work or heat currents (which are obtained from time-dependent contractions of elementary time-dependent currents). We emphasize that, being complementary to the results of Ref.\ \cite{Barato2018}, our bounds constitute by now the only known TURs for 
thermal devices with broken time reversal, see below. An interesting question is whether one can recover our LR results from a more general ansatz within the variational method of Ref.\ \cite{Barato2018}.

\subsection{B. Performance constraints}
Mesoscopic heat engines are generally characterized by two benchmark
parameters: power and efficiency \cite{Brandner2015},
\begin{equation}
P= -TF_w J_w \quad\text{and}\quad \eta\equiv P/J_q\leq 
\eta_{{{\rm C}}} = TF_q. 
\end{equation}
The Carnot bound $\eta_{{{\rm C}}}$ is thereby the only universal 
bound following from the laws of thermodynamics. 
For small driving amplitudes, further constraints on the two figures 
$P$ and $\eta$ can, however, be derived from the bound
\eqref{PT_Cl_BoundAI} on the asymmetry and the TUR Eq. (2). 

To this end, first, we notice that, in the linear response regime,
$\eta$ becomes
\begin{equation}
\eta = -\frac{TF_w(L_{ww}F_w + L_{wq}F_q)}{L_{qw}F_w +L_{qq}F_q}. 
\end{equation}
Maximizing these expressions with respect to $F_w$ and using 
\eqref{PT_Cl_BoundAI} yields the upper bound \cite{Brandner2013b}
\begin{equation}\label{PT_Cl_BndEta}
\eta\leq
\eta_{{{\rm C}}} x
\frac{\sqrt{4x\chi +(x-1)^2}-|x-1|}{\sqrt{4x\chi + (x-1)^2}+|x-1|}
\end{equation}
on the efficiency, where the dimensionless parameters 
\begin{equation}
0\leq \chi \equiv\frac{s_\L^2}{1+s_\L^2} \leq 1
\quad\text{and}\quad
x\equiv L_{wq}/L_{qw}
\end{equation}
describe the asymmetry of the Onsager coefficients \eqref{PT_Cl_OC}. 
Notably, \eqref{PT_Cl_BndEta} implies that, in the generic case 
$x\neq 1$, the Carnot $\eta_{{{\rm C}}}$ efficiency can be attained
only in the limit $\chi\rightarrow 1$, when \emph{asymmetry index $s_\L$ diverges}. For the case of periodically driven system, however, from~\eqref{PT_Cl_BoundAI} we have that $\chi$ describes the engine's deviation from ideal adiabatic operation,
\begin{equation}
 \chi \equiv \frac{\lambda^2}{\lambda^2 + \Omega^2},
\end{equation}
 so that $\chi\rightarrow 1$ corresponds to the adiabatic limit where $P$ vanishes.

The trade-off between power and efficiency can be captured 
quantitatively using the general TUR derived in the main text, see, 
also, \cite{Pietzonka2017b}. 
Specifically, by applying Eq. (2) of the main text, to the work flux $J_w$ and the 
heat uptake $J_q$, we find the relations 
\begin{equation}\label{PT_Cl_PEta1}
\frac{2T P}{D_w} \leq (\eta_{{{\rm C}}}/\eta-1) (1+s_\L^2)=\frac{\eta_{{{\rm C}}}/\eta-1}{1-\chi}
\end{equation}
and
\begin{equation}\label{PT_Cl_PEta2}
\frac{2T P}{D_q} \leq \eta(\eta_{{{\rm C}}}-\eta)(1+s_\L^2) =\frac{\eta(\eta_{{{\rm C}}}-\eta)}{1-\chi}.
\end{equation}
Here, $D_\alpha = 2 L_{\alpha\alpha}$ denotes the fluctuations of the
currents $J_\alpha$. The two bounds \eqref{PT_Cl_PEta1} and \eqref{PT_Cl_PEta2} show that an engine can reach Carnot 
efficiency only at the price of either: vanishing power, diverging
fluctuations in its output and input, or a diverging asymmetry index. For periodically driven engine operating in finite time, i.e., $\chi<1$, Carnot efficiency can only be achieved at vanishing power or with diverging fluctuations, cf.~\eqref{PT_Cl_BoundAI}. For further discussions see
\cite{Pietzonka2017b}.

\subsection{C. Example}
To illustrate the general theory outlined in this section, we consider
a paradigmatic and experimentally relevant system consisting of an
overdamped Brownian particle confined in a modulated harmonic trap~\cite{Schmiedl2008,Blickle2012,Martinez2015}.
In one dimension, the Hamiltonian of this engine can be written as
\begin{equation}
H_t(\xi) = \frac{\kappa_0}{2}\xi^2 + F_w f^w_t \frac{\kappa
_0}{4} \xi^2,
\end{equation} 
where $\xi$ denotes the position of the particle and $\kappa_0$ the
unperturbed strength of the trap. 
The free Fokker-Planck operator governing the dynamics of system
variables in equilibrium reads 
\begin{equation}
\mathsf{L}(\xi) = -\mu\kappa_0 \xi\partial_\xi + \mu T\partial_\xi^2, 
\end{equation}
where $\mu$ denotes the mobility of the particle. 
Choosing harmonic driving protocols, 
\begin{equation}
f^w_t = \cos(\Omega t) \quad\text{and}\quad
f^q_t = \cos(\Omega t +\phi),
\end{equation}
and evaluating \eqref{PT_Cl_OC} thus yields the Onsager matrix 
\begin{align}
\label{PT_Cl_HOOM}
&\mathbb{L}_{{{\rm HE}}} = \frac{\lambda(1-\chi)}{16}\\
&\times
\left(\!\begin{array}{cc}
1 & -2T \bigl[\cos\phi -\frac{\lambda}{\Omega}\sin\phi\bigr]\\
-2T \bigl[\cos\phi + \frac{\lambda}{\Omega}\sin\phi\bigr] & 4T^2
\end{array}\!\right)
\nonumber
\end{align}
with $\lambda\equiv 2\mu\kappa_0$ being the only relevant relaxation
rate of the system. 

It is straightforward to calculate the asymmetry index of the matrix
\eqref{PT_Cl_HOOM}, 
\begin{equation}
s_{{{\rm HE}}} = \lambda/\Omega,
\end{equation}
which saturates the bound \eqref{PT_Cl_BoundAI} irrespective of the 
phase shift $\phi$. 
Moreover, if the affinities are chosen such that
\begin{align}
F_w/F_q &= (\Omega/\lambda) \sin\phi - \cos\phi \quad\text{or}\\
F_w/F_q &= \frac{\lambda}{\Omega\sin\phi - \lambda\cos\phi}, \nonumber
\end{align}
the uncertainty products of $\mathcal{Q}_w$ and $\mathcal{Q}_q$ of 
work flux and heat uptake respectively attain their lower bound 
\begin{equation}
\mathcal{Q}_\alpha = \sigma D_\alpha/J_\alpha^2 \geq 
\frac{2}{1+s_{{{\rm HE}}}^2} = 2(1-\chi). 
\end{equation}
This result also entails that, at least in linear response, the 
harmonic Brownian heat engine discussed here can saturate both of the
power-efficiency trade-off relations, 
\eqref{PT_Cl_PEta1} and \eqref{PT_Cl_PEta2}, for suitably chosen affinities. 

\section{SM3: Bounds with positive semidefinite correlation matrix} \label{sec:var}

\subsection{A. Semidefinite correlation matrix}
 We further discuss the LR bounds Eqs. (6) and (8) of the main text
and the nonlinear bound Eq. (16) of the main text. 
We elaborate on the general case of positive semidefinite 
matrices $\D$ and $\L_S(\F)$ by considering the optimal current 
Eq. (14) of the main text, which was derived from the variational
principle for the relative uncertainty, see Eq. (14) of the main 
text. 

In LR, we have $\D=2\L_S$. 
Thus, for any $\F$ a solution of Eq. (14) of the main text exists 
only if the kernel of $\L_S$ lies in the orthogonal complement of 
the range of $\L_A$ such that the pseudoinverse $\L_S^{+}$ of $\L_S$
can be used to solve Eq. (14) of the main text. 
This condition is always met, for example, by Onsager matrices given
in terms of the Landauer-B\"uttiker formula, see Sec.~\ref{sec:MJinLR}
and~\cite{Brandner2013,Brandner2017}.

However, if the kernel of $\L_S$ overlaps with the range of $\L_A$,
for $\c$ chosen from their intersection, we have $\c^\T\L_S \c=0$, 
while there exist a choice of affinities such that 
$\c^\T\L\F=\c^\T \L_A \F \neq 0$. 
In this situation, there exist a nonvanishing purely reversible 
current with $0$ variance and thus infinite relative precision;
the asymmetry index of $\L$ is then also infinite, and the TUR Eq.~(8)
of the main text holds formally.

Analogously, beyond LR, a solution of Eq.~(14) of the main text
exists only if $\J$ is orthogonal to the kernel of $\D$. 
Otherwise, a purely reversible current with infinite relative 
precision exists and the RHS of Eq.~(16) of the main text diverges.

\subsection{B. Variational principle for optimal affinities in LR} 
We observe that, within LR, the variational principle Eq.~(14) of the
main text can be extended by including the affinities, that is, we 
have 
\begin{multline} \label{varC2}
\max_{\F}\max_{\c} J_\c^2/ (\sigma D_\c) =
\max_{\F}\max_{\c} \left(2 \c^{\T}\L\F -\sigma \c^\T \L_S \c\right)\\
=\max_{\F}\max_{\c} \left(2 \c^{\T} \L\F
-\F^\T \L_S \F \cdot \c^\T \L_S \c \right).
\end{multline}
Since the maximizations can be interchanged, we thus obtain a variational principle for an uncertainty of a fixed current, Eq.~(10) in the main text. It follows that the 
optimal affinity vector $\F_{{{\rm opt}}}$ for a given vector of
contraction coefficients $\c$, is determined by the condition 
\begin{equation}
D_{\c}\cdot\F_{\text{opt}}^{\T}\L_S = \c^{\T}\L,
\end{equation}
which, for the scale-invariant relative uncertainty 
$J_\c^2/ (\sigma D_\c)$, relaxes to
\begin{equation}
\F_{\text{opt}}^{\T}\L_S \propto \c^{\T}\L. 
\label{optF}
\end{equation}
corresponding to the condition~(11) in the main text. 
Analogously, the optimal vector of contraction coefficients, 
$\c_{{{\rm opt}}}$, for a given vector $\F$ must fulfil 
\begin{equation}
\sigma\cdot\L_S \c_{\text{opt}}= \L\F
\end{equation}
as a consequence of \eqref{varC2}. 
This condition is equivalent to Eq.~(15) of the main text up to
rescaling. 
Finally, we note that \eqref{varC2} is a variational principle for the
right-hand side of the bound Eq. (8) in the main text, and, thus, also
for the asymmetry index defined in Eq. (9) of the main text.

\section{SM4: Formal connection of the uncertainty product beyond linear response to the asymmetry index} \label{sec:var}

To establish a connection between the minimal uncertainty product for a current beyond linear response, Eq. (16) in the main text, and the TUR for the linear response, in Eqs. (6) and (8), we introduce, as a formal
generalization of the Onsager matrix $\L$, a nonequilibrium
conductance matrix $\L(\F)$ such that $\J=\L(\F)\F$ 
\footnote{Note that the nonequilibrium conductance matrix is itself
	a function of the affinities $\F$ and, in general, not uniquely 
	defined through the condition $\J=\L(\F)\F$.
	While, here, we consider $\L(\F)$ as a purely formal object, we note 
	that such a matrix can indeed be constructed in an unambiguous manner
	for time-homogeneous Markov processes using the framework of large
	deviation theory, for details see \cite{Vroylandt2018}.
}.
Dividing $\L(\F)$ into a symmetric and an antisymmetric part, 
$\L_S(\F)$ and $\L_A(\F)$, we thus obtain 
\begin{equation}\label{TURoptC2}
\frac{\J^\T  \D^{+} \J}{\F^\T\J}= \frac{\F^\T [\L_S^\T(\F)\, 
	\D^{+}\L_S(\F)+ \L_A^\T(\F)\, \D^{+}\L_A(\F)] \F}{\F^\T\L_S(\F)\,\F}. 
\end{equation}
Assuming further that $\L_S(\F)$ is invertible on $\J$, we arrive 
at the generalization of Eq.~(8) in the main text, 
\begin{eqnarray} 
&&\frac{J_\c^2}{\sigma D_\c}\leq \frac{\tilde{\F}^\T (\mathds{1}+\Y)
	\tilde{\F}}{2\,\tilde{\F}^\T\tilde{\F}}+\frac{\tilde{\F}^\T\X^\dagger
	(\mathds{1}+\Y)\X\tilde{\F}}{2\,\tilde{\F}^\T\tilde{\F}} 
\label{TURmaxF}\\
&&\leq \frac{1}{2}+ \max_{\tilde{\F}} \frac{1}{2}
\left[\frac{\tilde{\F}^\T \X^\dagger\X \tilde{\F}}{
	\tilde{\F}^\T\tilde{\F}}
+\frac{\tilde{\F}^\T(\mathds{1}+\X)^\dagger
	\Y(\mathds{1}+\X)\tilde{\F}}{\tilde{\F}^\T\tilde{\F}}\right].
\nonumber
\end{eqnarray}
Here, we have, similar to our LR analysis, 
$\tilde{\F}=[\L_S(\F)]^{1/2}\F$ and 
$\X=i[\L_S^+(\F)]^{1/2}\L_A(\F)[\L_S^+(\F)]^{1/2}$. 
Furthermore, the new matrix 
$\Y/2=\L_S^{1/2}(\F)[\D^+-\L_S^+(\F)/2]\L_S^{1/2}(\F)$ appears beyond
LR, since, in general, $\D\neq 2\L_S(\F)$.  
By optimizing the choice of affinities $\F$ in the second line of
\eqref{TURmaxF}, one formally arrives at a generalized TUR beyond 
linear response, cf. (8) and (9) in the main text.

\end{document}